\documentclass{elsart}

\usepackage{amssymb}

\def\S{{\mathcal S}}
\def\T{{\mathcal T}}
\def\U{{\mathcal U}}

\def\K{{\mathcal K}}
\def\G{{\mathcal G}}
\def\M{{\mathcal M}}
\def\bzero{\beta_0}

\def\CA{C_A}
\def\CF{C_F}

\def\TR{T_R}
\def\NF{N_F}
\def\as{\ensuremath{\alpha_{s}}}
\def\a0{\alpha_0}

\def\ket#1{|{#1}\rangle}
\def\bra#1{\langle{#1}|}

\def\cm{{\mathcal M}}

\def\MSbar{$\overline{{\rm MS}}$}

\def\bom#1{{\mbox{\boldmath $#1$}}}

\def \ep{\epsilon}

\def\npps#1#2#3{{ Nucl.~Phys.~Proc.~Suppl.}~{\bf #1} (#2) #3}
\def\npb#1#2#3{{ Nucl.~Phys.}~{\bf B#1} (#2) #3}
\def\plb#1#2#3{{ Phys.~Lett.}~{\bf B#1} (#2) #3}
\def\prd#1#2#3{{ Phys.~Rev.}~{\bf D#1} (#2) #3}
\def\jhp#1#2#3{{ JHEP}~{\bf #1} (#2) #3}
\def\zpc#1#2#3{{ Z.~Phys.}~{\bf C#1} (#2) #3}

\def\bea {\begin{eqnarray}}
\def\eea {\end{eqnarray}}

\def\be {\begin{equation}}
\def\ee {\end{equation}}

\begin{document}
\begin{frontmatter}
\begin{flushright}
YITP-SB-02-56
\end{flushright}

\title{Multi-loop Amplitudes and Resummation\thanksref{labelthanks}}
\thanks[labelthanks]{Partially supported by the National Science Foundation
   grant PHY-0098527}

\author{George Sterman} and
\ead{sterman@insti.physics.sunysb.edu}
\author{Maria E. Tejeda-Yeomans}
\ead{tejeda@insti.physics.sunysb.edu}
\address{
C.N.Yang Institute for Theoretical Physics \\
State University of New York \\
Stony Brook, NY 11794-3840 \\
}

\begin{abstract}
We explore the relation between resummation and explicit
multi-loop calculations for
QCD hard-scattering amplitudes. We describe how the
factorization properties of amplitudes lead to the
exponentiation of double and
single poles at each order of perturbation theory.
  For these amplitudes, previously-observed relations
between single and double poles in different $2\rightarrow 2$ processes can
now be interpreted in terms of universal functions associated with external
partons and process-dependent anomalous dimensions that describe coherent
soft radiation. Catani's proposal for multiple poles in
dimensionally-continued amplitudes emerges naturally.
\end{abstract}


\end{frontmatter}

\section{Introduction}
\label{sec:intro}

The past few years have seen a breakthrough in the calculation of two-loop
matrix elements and amplitudes in
QCD~\cite{alltwoloops,qqpg,bernamps,berngggg}. At the same time
significant progress has occured in the
resummation of logarithmic corrections to all orders in perturbation
theory~\cite{resrv,CLS,resum,deFlorian1}.
In this paper, we hope to contribute toward linking advances
in fixed-order QCD amplitudes and in resummation. We attempt to clarify how
the factorization properties of QCD hard scattering amplitudes
influence calculations at fixed order.

We will see how the single pole structure found at the level
of two loops in matrix elements
can be understood in terms of properties of a hitherto uncalculated two-loop
anomalous dimension for soft radiation. The knowledge of this two-loop soft
function is an important part of resummations at the level of
next-to-next-to-leading logarithms.
In addition, we will relate Catani's very successful 
proposal~\cite{catanitwo} for the
$\epsilon$-pole structure of dimensionally regularized ($D=4-2\ep$)
amplitudes at one and two loops to known factorization and resummation
properties of QCD amplitudes. These results will be illustrated for
amplitudes involving quarks and antiquarks.

\section{The factorized hard scattering amplitude}
\label{sec:facampsec}

Consider the partonic process
\begin{equation}
\label{partproc}
{\rm f}:\quad f_A(\ell_A,r_A) + f_B(\ell_B,r_B) \to f_1(p_1,r_1) +
f_2(p_2,r_2)\, ,
\end{equation}
which involves four partons with flavors $f_i$, momenta $\{ \ell_i, p_i \}$
and
color $r_i$. In perturbation theory, the hard scattering amplitude ${\M}^{\rm
   [f]}$
associated with this process can be written as a matrix in the space of the
color indices of its external partons $r_i=\{r_A, r_B; r_1, r_2\}$. It is
convenient to express these amplitudes in a basis of color tensors,
$\left(c_I\right)_{r_i}$, so that~\cite{KOS}
\begin{eqnarray}
\label{amp}
\M^{\rm[f]}_{\{r_i\}}\left(\{ \wp_j \},\frac{Q^2}{\mu^2},\as(\mu^2),\ep \right)
=
\M^{\rm[f]}_{L}\left( \{\wp_j\},\frac{Q^2}{\mu^2},\as(\mu^2),\ep \right)
\, \left(c_L\right)_{\{r_i\}}\,.
\end{eqnarray}
Here $i,j=A,B,1,2$ and $Q^2$ is the overall hard-scattering scale, which
we may choose as $s$ for $2\rightarrow 2$ scattering. The vectors
$\wp_j=p_j/\sqrt{s/2}$, specify the directions of the incoming and outgoing
lines and hence the full kinematics of the process\footnote{For more than
   two particles in the final state, we must include information on the
   relative energies of the outgoing lines.}.
After renormalization, these amplitudes retain poles in
$\ep$, due to infrared and collinear singularities.
The singular behavior of the $2 \rightarrow 2$ amplitude is entirely due
to internal lines that carry momenta that are either
soft or collinear to one of the external lines.  For gauge-invariant
sets of diagrams, these singularities are entirely logarithmic
in four dimensions, corresponding to poles at $\ep=0$.

The amplitude $\M^{\rm[f]}$ for partonic process f can be
factorized into products of functions that
organize the contributions of momentum regions relevant
to $\ep$ poles in the scattering amplitude~\cite{Sen,Akh}.
These are: i) process-dependent functions $h^{\rm[f]}_{I}$ that describe the
short-distance dynamics of the hard scattering, one function for
each of the elements of the basis of color exchange;  ii) a matrix of
functions, $S^{\rm[f]}_{LI}$ that describes the coherent soft radiation
arising from the overall color flow and  iii) a function $J^{\rm[f]}$,
dependent only on the list of external partons, and otherwise
independent of the color flow, which describes the perturbative evolution of
the incoming and outgoing partons for flavors $f_i=q,\bar q,g$. In
these terms, the factorized form of the amplitude associated with the
process ${\rm f}$ of Eq.~(\ref{partproc}) is~\cite{Sen,KOS}
\begin{eqnarray}
\label{facamp}
\M^{\rm[f]}_{L}\left(\wp_i,\frac{Q^2}{\mu^2},\as(\mu^2),\ep \right)
&=&J^{\rm[f]}\left(\frac{Q^2}{\mu^2},\as(\mu^2),\ep \right)
S^{\rm[f]}_{LI}\left(\wp_i,\frac{Q^2}{\mu^2},\as(\mu^2),\ep \right) \nonumber
\\
&&\times ~~h^{\rm[f]}_{I}\left(\wp_i,\frac{Q^2}{\mu^2},\as(\mu^2)
\right)\, ,
\end{eqnarray}
where we collect the various virtual "jet" factors associated with external
partons in
\begin{eqnarray}
\label{psidesc}
J^{\rm[f]}\left(\frac{Q^2}{\mu^2},\as(\mu^2),\ep \right)
&\equiv &
\prod_{i=A,B,1,2}J^{[f_i]}_{\rm{(virt)}}\left(\frac{Q^2}{\mu^2},\as(\mu^2),\ep
\right) \, .
\end{eqnarray}
The products are over the two incoming lines, and over the two outgoing
lines (or more, in the cases where we generalize the $2\rightarrow 2$ process
of Eq.~(\ref{partproc})). Notice that in Eq.~(\ref{facamp}) only the soft and
hard-scattering functions depend on the momenta $\wp_i$\footnote{To shorten
  the notation, $S_{LI}$ and $h_{I}$ dependence on $\wp_i$ , will be
  implicit.}.  
The parameter $\mu$ remains the renormalization scale while the factorization
scale is chosen as the hard scale $Q$.

The predictive power of the factorized form of the amplitude follows from the
properties of its individual functions. The jet functions in
  $J^{\rm[f]}$ include all collinear dynamics, and hence all double
poles in dimensional regularization, which arise from the overlap of infrared
and collinear enhancements in perturbation theory. The matrix of soft
functions $S^{\rm[f]}_{LI}$ provides at most a single infrared pole per loop
and, although it depends on the process kinematics it is otherwise completely
determined by a set of anomalous dimensions that can be computed in
the eikonal approximation.  The hard-scattering functions $h^{\rm[f]}_{I}$ are
fully infrared finite.  The jet functions and the soft function
can be defined in terms of specific QCD matrix elements~\cite{KOS,CTG}.

The independence of the amplitude $\M$ from the choice of the factorization
scale $Q$ leads, by the usual connection between factorization and
evolution, to the renormalization group equation~\cite{CLS}
\begin{eqnarray}
\label{muS}
{d\over d\ln\, Q}\, S_{LI}  &=& - \Gamma^{\rm[f]}_{LJ}\,\, S_{JI},
\end{eqnarray}
where the variations of the jet functions and the soft matrix with the
scale $Q$ are compensated by variations of the hard function.

There is still considerable freedom in the construction of the
jet and soft functions.
For examples, we may shift infrared finite contributions between
the jet and hard scattering functions, and/or single-logarithmic,
process-independent soft contributions between the soft matrix
and the jets.  Specifically, the soft matrix $S_{LI}$
is defined only up to any multiple of the identity matrix in
the color basis space.
Below, we will identify a convenient, but by no means unique, scheme for
the definition of the various functions in (\ref{facamp}).

Of primary importance here is the observation that mixing of color structures
as a result of soft parton exchange is entirely contained within the matrix
$\Gamma^{\rm [f]}$, and is therefore summarized by the following solution to
the  renormalization group for the soft function,
\begin{eqnarray}
\label{expoS}
{\bf S}^{\rm[f]}\left(\frac{Q^2}{\mu^2},\as(\mu^2),\ep \right) =
{\rm P}~{\rm exp}\left[ -\frac{1}{2}\int_{0}^{-Q^2} \frac{d\tilde{\mu}^2}{\tilde{\mu}^2}
{\bf \Gamma}^{\rm[f]} \left(\bar\as\left(\frac{\mu^2}{\tilde{\mu}^2},\as(\mu^2),\ep\right)\right) \right],
\end{eqnarray}
where P stands for path ordering. In the dimensionally-regularized theory,
the effective coupling should be thought of as expanded in powers of
$\as(\mu^2)$. To the accuracy we work, this is given by the one-loop 
form~\cite{magnea}
\begin{eqnarray}
\label{asinD}
\bar\as\left(\frac{\mu^2}{\tilde{\mu}^2},\as(\mu^2),\ep\right) &=&
\as(\mu^2)\left(\frac{\mu^2}{\tilde{\mu}^2} \right)^\ep
\sum_{n=0}^{\infty}\left[\frac{\bzero}{4\pi\ep}
\left(\left(\frac{\mu^2}{\tilde{\mu}^2} \right)^\ep-1\right) \as(\mu^2)
\right]^n
\end{eqnarray}
where
$$\bzero = \frac{11}{3}\CA - \frac{2}{3}\NF.$$ The integrals arising in
the exponential of Eq.~(\ref{expoS}) are quite simple and for comparison to
fixed $n$th-order
calculations, we only need to collect all contributions up to
${\mathcal O}\left({\as^n(\mu^2)}\right)$.
All the one-loop anomalous dimensions in these equations have been computed
previously in Refs.~\cite{KOS}, so that the color mixing due to
single soft gluon exchange can be predicted (and checked) for specific
processes.

In the next section we will use a specific
definition of the jet function, based on factorization and on the
application of Eq.\ (\ref{facamp}) to the simplest color flow of all, the
time-like Sudakov form factor.

\section{Jet and soft functions}

A convenient explicit expression for jet functions may be developed by
applying the factorization in Eq.\ (\ref{facamp}) to the singlet form
factors for quarks and gluons, which we denote by $\M^{[i \bar i \to 1]}$. To
be specific, we consider the quark Sudakov form factor. In this case,
there is no color mixing at all, and the soft anomalous dimension reduces to
a number. In the following, we shall absorb this number into the 
evolution of the jets,
and reduce the soft function to unity {\it by definition}. Similarly, given
that the anomalous dimension for any electroweak vertex vanishes in QCD, we
   may also reduce the hard function to unity in this case.

We thus define our jet functions by the relation
\begin{eqnarray}
\label{jsudakov}
J^{[i]}\left(\frac{Q^2}{\mu^2},\as(\mu^2),\ep\right)
  = J^{[\bar i]}\left(\frac{Q^2}{\mu^2},\as(\mu^2),\ep\right)
&=& \left[\M^{[i \bar i \to
  1]}\left(\frac{Q^2}{\mu^2},\as(\mu^2),\ep\right)\right]^{\frac{1}{2}}
  \nonumber \\
\end{eqnarray}
where $Q^2$ is the characteristic momentum transfer, $\mu$ is the \MSbar
~renormalization scale ($\mu^2 = \mu_0^2 {\rm exp}[-\ep(\gamma_E-{\rm
   ln}4\pi)]$) and the electromagnetic Sudakov form factor in $D=4-2\ep$
dimensions is given by\footnote{A similar definition may be given for gluon
  jets in terms of matrix elements of conserved, singlet
  operators.}\cite{magnea,sudakovCDR}
\begin{eqnarray}
\label{solutionev}
\M^{[i \bar i \to 1]}\left(\frac{Q^2}{\mu^2},\as(\mu^2),\ep\right)
&&
={\rm exp}\Biggl\{
~\frac{1}{2} \int_{0}^{-Q^2}\frac{d\xi^2}{\xi^2}
\Biggl[
\K^{[i]}(\as(\mu^2),\ep)
\nonumber \\
&&~~~~~~~~~~~+\G^{[i]}\left(-1,\bar\as\left(\frac{\mu^2}{\xi^2},\as(\mu^2),\ep,\right)\ep\right)
\nonumber \\
&&~~~~~~~~~~~+\frac{1}{2}\int_{\xi^2}^{\mu^2} 
\frac{d\tilde{\mu}^2}{\tilde{\mu}^2}
\gamma^{[i]}_{K}\left(\bar\as\left(\frac{\mu^2}{\tilde{\mu}^2},\as(\mu^2),\ep
  \right)\right)
\Biggr] ~\Biggr\}. \nonumber \\
\end{eqnarray}

In Eq.~(\ref{solutionev}) the $\xi$ integration is defined order by 
order in perturbation theory
and as in our discussion of the soft function, the running
coupling $\bar \as$ is thought of as being expanded in powers of
$\as(\mu^2)$. As described in~\cite{sudakovCDR,evolsuda}, the functions
$\gamma^{[i]}_{K}$, $\K^{\rm [f]}$ and $\G^{[i]}$ may be read off by
comparison to explicit fixed-order
calculations~\cite{suda1L,suda2L}. In particular, $\K^{[i]}$ is
defined as a series of poles in $\ep$, while
$\G^{[i]}(-1,\as,\ep)$ includes the non-singular $\ep$-dependence before integration. Notice
that in Eq.~(\ref{solutionev}) all the $Q$ dependence of the 
Sudakov form factor is organized by the dimensionally-regulated running coupling and that by
   construction $\K^{[q]}=\K^{[\bar q]}$, $\G^{[q]}=\G^{[\bar q]}$ and 
$\gamma^{[q]}_K=\gamma^{[\bar q]}_K$~\cite{sudakovCDR,factorjets}.

Following the method of Refs.~\cite{sudakovCDR}, we verify that
$\gamma^{\rm [i]}_{K}$ is the familiar Sudakov double-logarithmic
anomalous dimension known up to two loops~\cite{gammagq} to be
\begin{equation}
\label{gammaKexpand}
\gamma^{\rm [i]}_{K}= 2~C_{i}\left({\as\over \pi}\right)\, \left[ 1+
\left({\as\over \pi}\right)\frac{K}{2} \right]
\end{equation}
with $C_q=C_F,\ C_g=C_A$ and $K = C_A \left(67/18-\zeta(2))-N_F(5/9)\right)$.

Before going on to elastic scattering amplitudes, we
observe that our choice of jet function~(\ref{jsudakov})
corresponds to
a particular scheme for the factorization of the Sudakov
amplitude in which the hard-scattering and soft functions
are set to unity, $h^{[i\bar i \to 1]}= 1$ in Eq.~(\ref{muS})
for the form factor. In this scheme, the hard scattering
and soft functions of $2\rightarrow 2$ amplitudes
will be computed with these ``Sudakov-defined" jet
functions.  In particular, the
matrices $\Gamma^{\rm[f]}$ found in this way
correspond to those calculated in Refs.~\cite{KOS,CTG} at one loop. We will
return to the
issue of alternative choices below.

\section{Pole structure for multi-loop hard-scattering amplitudes }

The combination of Eqs.~(\ref{expoS}) and (\ref{jsudakov}) for the soft
and jet functions applied to Eq.~(\ref{facamp}), allow us to give a useful
expression for an arbitrary hard-scattering amplitude at two loops and
beyond.  In particular, the all-orders structure
of any $2\rightarrow n$ amplitiude at wide angles is readily understood
as the exponentiation of an appropriate
power of the singlet form factors times the expansion of
the exponentiated soft anomalous dimension.  With input to
$n+1$ loops in the singlet form factors of quark and gluon,
and $n$ loops in the soft anomalous dimension, we
can predict the full n$th$-next-to-leading poles in
dimensional regularized amplitudes, in much the same way we predict powers
of logarithms in threshold or other resummations.

In particular, on the basis of two-loop calculations for the jet functions
and of the Sudakov form factor, we can predict almost the entire 
two-loop single-pole term for
explicit calculations involving quarks.
Conversely, on the basis of two-loop calculations
we can also determine the soft anomalous dimensions at this order, 
and proceed to
a full next-to-next-to-leading pole approximation.

For ease of comparison, we follow
Refs.~\cite{catanitwo} and denote the vector in
color space $\M_{r_i}$ with components $\M_L$ in Eq.~(\ref{amp}) projected
into the orthogonal basis $\{c_i\}$ as
\begin{eqnarray}
\label{notationM}
\ket{\cm^{\rm[f]}} \equiv \M^{\rm[f]}_{L}\left( 
\{\wp_j\},\frac{Q^2}{\mu^2},\as,\ep \right)\,
\left(c_L\right)_{\{r_i\}}
\end{eqnarray}
whose perturbative expansion is
\begin{eqnarray}
\label{mamp}
\ket{\cm^{\rm[f]}} &=& \sum_{m=0}^{\infty}
\left(\frac{\as(\mu^2)}{\pi}\right)^m
\ket{\cm^{[{\rm f}(m)]}}.
\end{eqnarray}

Applying Eqs.~(\ref{expoS}) -- (\ref{solutionev}) to the factorized form of the
amplitude in Eq.~(\ref{facamp}), we easily collect all the singular structure
of the one-loop amplitude, using $\K^{[{\rm f}(1)]}=\gamma^{[{\rm
    f}(1)]}_K/(2\ep)$, into a function ${\bom F}^{[{\rm f}(1)]}$
as\footnote{In the following, we use the subscript {\it fin} to denote finite
  functions and the subscript {\it ren} to denote renormalized functions.}
\begin{eqnarray}
\label{1Lamp}
\ket{\cm^{[{\rm f}(1)]}_{ren}} &=& {\bom F}^{[{\rm f}(1)]}(\ep) 
\ket{\cm^{[{\rm f}(0)]}} +
\ket{\cm^{[{\rm f}(1)]}_{fin}},
\end{eqnarray}
with the lowest order amplitude
\begin{eqnarray}
\label{bornamp}
\ket{\cm^{[{\rm f}(0)]}} &=& \ket{\cm^{[{\rm Born}]}}.
\end{eqnarray}

The function ${\bom F}^{[{\rm f}(1)]}(\ep)$ in (\ref{1Lamp}) is given
in terms of the one-loop coefficients in Eqs.\ (\ref{expoS}) and 
(\ref{solutionev}) by
\begin{eqnarray}
\label{f1e}
{\bom F}^{[{\rm f}(1)]}(\ep) & \equiv &
\frac{1}{2}\Biggl[ - \left(\frac{\gamma^{[{\rm f}(1)]}_K}{2\ep^2} +
     \frac{\G^{[{\rm f}(1)]}_0}{\ep}\right){\bf 1}
+ \frac{{\bf \Gamma}^{[{\rm f}(1)]}}{\ep}
\Biggr]\left(-\frac{\mu^2}{Q^2} \right)^{\ep}.
\end{eqnarray}
Here, we define
\begin{eqnarray}
\G^{\rm [f]} \equiv ~\frac{1}{2}\sum_{i\in {\rm f}} \G^{[i]}, \quad \quad
\gamma^{\rm [f]}_{K} \equiv \frac{1}{2}\sum_{i\in {\rm f}} 
~\gamma^{[i]}_{K}\, ,
\end{eqnarray}
and $\G^{[{\rm f}(1)]}_0 \equiv \G(Q^2/\mu^2=1,\ep=0)$.
The $\ep$-poles at ${\mathcal O}\left({\as}\right)$ are organized by this divergent function into
those that are color uncorrelated, which are given by Sudakov
exponentiation through $\G$ and $\gamma_K$, and those that are color
correlated, which are collected by the one-loop soft anomalous dimension
matrix $\Gamma^{[{\rm f}(1)]}$.

The finite reminder in Eq.\ (\ref{1Lamp}) is the one-loop
hard scattering function of Eq.\ (\ref{facamp}), in the notation,
\be
\ket{\cm^{[{\rm f}(1)]}_{fin}} = 
h^{[{\rm f}(1)]}_{I}\left(\wp_i,\frac{Q^2}{\mu^2} \right)\,
\left(c_I\right)_{\{r_i\}}
\ee

At two loops, we can conveniently organize the expansion
of Eq.\ (\ref{facamp}) by making use of the
divergent structure identified at one loop through ${\bom F}^{[{\rm 
f}(1)]}$, and by taking into
account that $\K^{[{\rm f}(2)]}=-\bzero\gamma^{[{\rm f}(1)]}_K/(16\ep^2) +
\gamma^{[{\rm f}(2)]}_K/(4\ep)$ and $\gamma^{[{\rm f}(2)]}_K = \frac{\rm
  K}{2}\gamma^{[{\rm f}(1)]}_K$. We find
\begin{eqnarray}
\label{2Lamp}
\ket{\cm^{[{\rm f}(2)]}_{ren}}
&=& {\bom F}^{[{\rm f}(1)]}(\ep) \ket{\cm^{[{\rm f}(1)]}}
+ {\bom F}^{[{\rm f}(2)]}(\ep) \ket{\cm^{[{\rm f}(0)]}}
+ \ket{\cm^{[{\rm f}(2)]}_{fin}} \nonumber \\
\end{eqnarray}
where
\begin{eqnarray}
\label{f2e}
{\bom F}^{[{\rm f}(2)]}(\ep) \equiv &&
-\frac{1}{2}\left[{\bom F}^{[{\rm f}(1)]}(\ep)\right]^2 + 
\frac{1}{2}\left({\rm K}
+ \frac{\bzero}{2\ep} \right){\bom F}^{[{\rm f}(1)]}(2\ep)\nonumber\\
&\ & \quad \quad
- \frac{\bzero}{4\ep}{\bom F}^{[{\rm f}(1)]}(\ep)
+\frac{1}{2}{\bom L}^{[{\rm f}(2)]}(2\ep)
\end{eqnarray}
with
\begin{eqnarray}
\label{h2funcb}
{\bom L}^{[{\rm f}(2)]}(\ep) \equiv && \frac{1}{\ep}\Biggl[
- \left( \G^{[{\rm f}(2)]}_0 - \frac{{\rm K}}{2} \G^{[{\rm f}(1)]}_0 
\right){\bf 1}
+{\bf \Gamma}^{[{\rm f}(2)]}-\frac{{\rm K}}{2}{\bf
   \Gamma}^{[{\rm f}(1)]} ~\Biggr]\left(-\frac{\mu^2}{Q^2}
\right)^{\ep}\, .
\end{eqnarray}

To evaluate the expressions given above, we use
\begin{eqnarray}
\label{g1fin}
\G^{[{\rm q}(1)]}_{0}&\equiv & \G^{[{\rm
     q}(1)]}\left(Q^2/\mu^2=1,\ep=0\right) \nonumber \\
&=& \CF \frac{3}{2},\\
\label{g2fin}
\G^{[{\rm q}(2)]}_{0}&\equiv & \G^{[{\rm
     q}(2)]}\left(Q^2/\mu^2=1,\ep=0\right) \nonumber \\
&=& \CF~\Biggl\{
3\left( \frac{1}{16} -\frac{1}{2}\zeta{(2)}+\zeta{(3)} \right)\CF
+\left(\frac{2545}{108} + \frac{11}{3}\zeta{(2)} - 13 \zeta{(3)}
     \right)\frac{\CA}{4} \nonumber \\
&&~~-\left(\frac{209}{108} + \frac{1}{3}\zeta{(2)} \right)~\TR \NF \Biggr\},
\end{eqnarray}
which are found by direct comparison of Eq.~(\ref{solutionev}) with the known
fixed-order results~\cite{sudakovCDR,suda1L,suda2L} for the quark form factor.

 From Eqs.~(\ref{1Lamp}) -- (\ref{f1e}), we see that the bulk of the one-loop
singular behavior is determined by a combination of
the Sudakov amplitudes $\M^{[i \bar i \to 1]}$, from which we can find
$\gamma_K^{[{\rm f}]}$ and $\G^{[{\rm f}]}$, and the one-loop anomalous
dimension matrix $\Gamma^{[{\rm f}(1)]}$.
Together, these functions control the two-loop singular behavior
from $\ep^{-4}$ down to $\ep^{-2}$.
Information specific to a $2\rightarrow n$ process at this order appears
first at ${\mathcal O}\left({\ep^{-1}}\right)$ in the contribution of the two-loop anomalous dimension
matrix $\Gamma^{[{\rm f}(2)]}$, which can be determined by comparison to an
explicit calculation. We may verify that the expressions above coincide with
the proposal of Catani for the structure of two-loop hard scattering
amplitudes, by direct comparison with Ref.~\cite{catanitwo}.

As an application, consider the
explicit two-loop calculation for the
process $q\bar{q}\rightarrow q\bar{q}$~\cite{qqpg}.
To express the amplitude as in Eq.\ (\ref{amp}),
we employ the basis of $t$-channel singlet and
octet color exchange (with $T^a$ the group generators in the fundamental
representation)~\cite{qqpg,KOS,CTG},
\begin{eqnarray}
\label{c1c2}
c_1 =\delta_{r_1r_A}\; \delta_{r_2r_B}\, &,& c_2 = \sum_a T^a_{r_1r_A}\;
T^a_{r_2r_B}\, .
\end{eqnarray}
In this basis the one-loop soft anomalous dimension matrix\footnote{It
   corresponds to the one presented in
   Eq.(51) of Ref.~\cite{KOS}, when we choose the scale to be
   the $\mu^2=\hat{s}=(\ell_A + \ell_B)^2$.} is given by
\begin{eqnarray}
\label{Gamma1Lex}
{\bf \Gamma}^{[{\rm q}(1)]} &=& \left[
\begin{array}{ccc}
2\CF \T &~ & \frac{\CF}{N}\left(\S - \U \right)   \\
2\left(\S - \U \right) &~ &\frac{N^2-2}{N}\S - \frac{1}{N}\left(\T - 
2\U \right)
\end{array}
\right]
\end{eqnarray}
where
\begin{eqnarray}
\S \equiv {\rm ln}\left( -\frac{s}{\mu^2} \right),
~~\T \equiv {\rm ln}\left( -\frac{t}{\mu^2} \right),
~~\U \equiv {\rm ln}\left( -\frac{u}{\mu^2} \right)
\end{eqnarray}
and $\{s, t, u\}$ are the usual Mandelstam variables.

If we use Eq.~(\ref{Gamma1Lex}) and apply the explicit expressions 
for $\gamma^{[{\rm q}(1)]}_K$,
$\gamma^{[{\rm q}(2)]}_K$ and $\G^{[{\rm q}(1)]}_0$ given
above, we derive predictions for the poles $\ep^{-4}$, $\ep^{-3}$
and $\ep^{-2}$ that are in complete agreement with the explicit 
calculation of~\cite{qqpg}.
We can therefore use the $\ep^{-1}$ term of this calculation
to determine the contribution of $\Gamma^{[{\rm f}(2)]}$
for this process.
More precisely, we can determine its
color-uncorrelated part, since this is what is
given in~\cite{qqpg}.

This direct comparison of the color
uncorrelated amplitudes found
  Refs.~\cite{qqpg} for $q \bar q$ scattering
matrix-elements by diagrammatic evaluation gives the
following surprisingly simple result
\begin{eqnarray}
\label{compareH}
\bra{\cm^{(0)}}\left({\bf \Gamma}^{\rm(2)}_{S}-\frac{{\rm K}}{2}{\bf
   \Gamma}^{\rm(1)}_{S}\right)\ket{\cm^{(0)}} =
   \CF\bzero\left(\frac{\zeta(2)}{16} + 1 \right) \bra{\cm^{(0)}} {\bf 1}
\ket{\cm^{(0)}}\, .
\end{eqnarray} 
Thus, comparison of Eqs.~(\ref{2Lamp}) -- (\ref{h2funcb}) with the explicit
calculation shows that most of the pole terms at two loops are
understandable through exponentiation. The remaining coherent part, sensitive
to the details of the process, is much simplified once exponentiation is
taken into account.

Even this remaining part is subject to further simplification,
as is strongly suggested by the presence of the overall
factor $\beta_0$, on the right of Eq.\ (\ref{compareH}).
In the scheme we have used above, the Sudakov form factor
directly determines the jet definition.
As we have pointed out, this is equivalent to setting $h^{[i\bar i \to 1]}=1$.
It is easy to show that choosing a nonzero $h^{[i\bar i \to 1(1)]}$ for
$\M^{[i\bar i \to 1(1)]}$ produces a corresponding change in $\G^{\rm (f,2)}$
proportional to $\beta_0 \, h^{[i\bar i \to 1(1)]}$. The importance of the
analogous scheme dependence in $k_T$-resummation has been emphasized 
recently in
Ref.~\cite{deFlorian1}. Thus, by using our freedom to modify
the hard-scattering function in the Sudakov
form factor, or equivalently to change the normalization
of the jet function, we can easily impose the condition
\begin{eqnarray}
\label{absorb0}
\bra{\cm^{(0)}}\left({\bf \Gamma}^{\rm(2)}_{S}-\frac{{\rm K}}{2}{\bf
   \Gamma}^{\rm(1)}_{S}\right)\ket{\cm^{(0)}} = 0\, .
\end{eqnarray}

Given our ability to impose Eq.~(\ref{absorb0}),
the uncorrelated part of the single poles
can be absorbed entirely into appropriately-chosen
jet functions. This freedom explains the
relations found in Refs.~\cite{qqpg,berngggg}. There, on the basis of direct
calculations it was found that for any $2 \rightarrow 2$ QCD process ${\rm
   f}$ with $n_g$ external gluons and $n_{q}$ external quarks and antiquarks
\begin{eqnarray}
\label{h2rel}
H^{[{\rm f}(2)]} = n_q H^{(2)}_q + n_g H^{(2)}_g,
\end{eqnarray}
where $H^{[{\rm f}(2)]}$ is the coefficient of the identity in ${\bom
   L}^{[{\rm f}(2)]}(\ep)$ on Eq.~(\ref{h2funcb}).

To determine the two-loop anomalous dimension fully, we will need
two-loop amplitudes as vectors in the basis states, not only
in uncorrelated form. These have been given so far for gluon-gluon
scattering~\cite{berngggg}, with an expression for single-pole
terms that is consistent with our results.

\section{Higher orders}

It is clear that the pattern described above extends to higher orders.
For example, the three-loop hard scattering amplitude has the structure
\begin{eqnarray}
\hspace{-1cm} \ket{\cm^{[{\rm f}(3)]}_{ren}} =
 {\bom F}^{[{\rm f}(1)]}(\ep) \ket{\cm^{[{\rm f}(2)]}}
+ {\bom F}^{[{\rm f}(2)]}(\ep) \ket{\cm^{[{\rm f}(1)]}}
+ {\bom F}^{[{\rm f}(3)]}(\ep) \ket{\cm^{[{\rm f}(0)]}}
+ \ket{\cm^{[{\rm f}(3)]}_{fin}} \nonumber \\ \nonumber \\
\end{eqnarray}
where
\begin{eqnarray}
{\bom F}^{[{\rm f}(3)]}(\ep) &=&
-\frac{1}{3}\left[{\bom F}^{[{\rm f}(1)]}(\ep)\right]^3
- \frac{1}{3}{\bom F}^{[{\rm f}(1)]}(\ep){\bom F}^{[{\rm f}(2)]}(\ep)
- \frac{2}{3}{\bom F}^{[{\rm f}(2)]}(\ep){\bom F}^{[{\rm f}(1)]}(\ep)
\nonumber \\
&&-\left(\frac{\bzero}{4\ep} \right)^2{\bom F}^{[{\rm f}(1)]}(3\ep)
+ \left(\frac{\bzero}{4\ep}\right)
\Biggl\{
-\frac{1}{2}\left[{\bom F}^{[{\rm f}(1)]}(\ep)\right]^2 -{\bom 
F}^{[{\rm f}(2)]}(\ep) \nonumber \\
&&~~~~~~~~~~+\frac{1}{2}\left({\rm K} + \frac{\bzero}{2\ep}\right)\left[2{\bom
     F}^{[{\rm f}(1)]}(3\ep) - {\bom F}^{[{\rm f}(1)]}(2\ep) \right] 
\nonumber \\
&&~~~~~~~~~~+{\bom L}^{[{\rm f}(2)]}(3\ep) 
- \frac{1}{2}{\bom L}^{[{\rm f}(2)]}(2\ep)\Biggr\}
+ \frac{1}{2}{\bom L}^{[{\rm f}(3)]}(3\ep),
\end{eqnarray}
and
\begin{eqnarray}
{\bom L}^{[{\rm f}(3)]}(\ep) \equiv &&
\Biggl[ - \left(\frac{\gamma^{[{\rm f}(3)]}_K}{2\ep^2} +
     \frac{\G^{[{\rm f}(3)]}_0}{\ep}\right){\bf 1}
+ \frac{{\bf \Gamma }^{[{\rm f}(3)]}}{\ep}
\Biggr]\left(-\frac{\mu^2}{Q^2} \right)^{\ep}.
\end{eqnarray}
These expressions enable us to predict the poles in $\ep$ from sixth order
down to fourth on the basis of the two-loop Sudakov form factor
and the one-loop soft anomalous dimension matrices.  Beyond this level,
the three-loop\footnote{The fermionic contributions to 
$\gamma_{K}^{[{\rm f}(3)]}$
have been calculated recently {\protect~\cite{a3}}} 
$\gamma_{K}^{[{\rm f}(3)]}$ and two-loop
soft anomalous dimension matrices will control all pole structure 
down to double poles,
while the color correlations in the single poles will determine ${\bf 
\Gamma}^{[{\rm f}(3)]}$.

\section{Conclusions and outlook}

We have shown how the factorization properties of
hard-scattering amplitudes enable us to understand
many of the complexities of their infrared
poles in dimensional regularization. In particular,
the universality of exponentiated collinear and soft singularities
associated with incoming and outgoing partons
provides a way of understanding important regularities predicted for
and found in explicit two-loop calculations.  They also
predict the structure, and for the leading terms the coefficients,
of poles to any order.

The program of calculating two-loop dimensionally-regulated amplitudes is
a step toward standard model hard-scattering cross sections beyond 
next-to-leading
order.  In this context, it is useful to note that
the factorization in Eq.\ (\ref{facamp}) for amplitudes
is shared by cross sections in the limit of partonic threshold,
where the parton center-of-mass energy is just
large enough to produce the final state.
At any fixed order in perturbation theory, leading
collinear and infrared singularities occur at partonic
threshold.  Therefore, the exponentiated structure
of singularities in $\ep$ for amplitudes will
reoccur in the singularities of cross sections with
radiation.
This observation is the basis
for threshold resummation, which may be a useful guide
in organizing the very challenging phase space integrals
in next-to-next-to-leading order cross sections.

Here we have explored singularities associated with purely virtual
corrections, but the simplifications that we have
found suggest that a similar analysis may be
useful for inelastic processes.
Leading singularities in cross sections cancel between
sets of cut diagrams that include the elastic amplitudes
at two loops.  The structure of singularities
in these amplitudes, as organized above,
  may help streamline the calculation of cross sections at two loops and
  beyond.

\begin{ack}
This work is supported in part by the National Science Foundation
   grant PHY-0098527. We would like to thank C. Anastasiou, C. F. Berger,
   T. Gehrmann, E. W. N. Glover, A. Kulesza and S. Moch for helpful
   discussions. Thanks also to Z. Bern and L. Dixon for critical
readings of the original version.
\end{ack}


\begin{thebibliography}{00}
\bibitem{alltwoloops} J. B. Tausk, \plb{469}{1999}{225}
   [arXiv:hep-ph/9909506]; V. A. Smirnov \plb{460}{1999}{367} 
[arXiv:hep-ph/9905323]
C. Anastasiou, E. W. N. Glover, C.  Oleari and
    M. E. Tejeda-Yeomans, \npb{605}{2001}{486} [arXiv:hep-ph/0101304];
E. W. N. Glover and M. E. Tejeda-Yeomans, \npb{605}{2001}{467}
[arXiv:hep-ph/0102201]; E. W. N. Glover and M. E. Tejeda-Yeomans,
\jhp{05}{2001}{010} [arXiv:hep-ph/0104178];
T. Binoth, E. W. N. Glover, P. Marquard and J. J. van der Bij,
\jhp{05}{2002}{060} [arXiv:hep-ph/0202266]; L. W. Garland, T. Gehrmann,
E. W. N. Glover, A. Koukoutsakis and E. Remiddi, \npb{642}{2002}{227}
[arXiv:hep-ph/0206067]
\bibitem{qqpg} C. Anastasiou, E. W. N. Glover and M. E. Tejeda-Yeomans,
   \npb{629}{2002}{255} [arXiv:hep-ph/0201274]; C. Anastasiou,
   E. W. N. Glover, C.  Oleari and M. E. Tejeda-Yeomans, \npb{601}{2001}{318}
   [arXiv:hep-ph/0010212]; \npb{601}{2001}{341} [arXiv:hep-ph/0011094] {\it
     ibid} \plb{506}{2001}{59} [arXiv:hep-ph/0012007]
\bibitem{bernamps} Z. Bern, A. De Freitas and L. Dixon, \jhp{11}{2001}{031}
   [arXiv:hep-ph/0109079]; Z. Bern, A. De Freitas and L. Dixon, 
\jhp{09}{2001}{037}
   [arXiv:hep-ph/0109078]
\bibitem{berngggg} Z. Bern, A. De Freitas and L. Dixon, \jhp{03}{2002}{018}
    [arXiv:hep-ph/0201161]
\bibitem{resrv} N.\ Kidonakis, Int.\ J.\ Mod.\ Phys.\ {\bf A15} (2000) 1245
[arXiv:hep-ph/9902484].
\bibitem{CLS} H. Contopanagos, E. Laenen and G. Sterman,
   \npb{484}{1997}{303} [arXiv:hep-ph/9604313]
\bibitem{resum} E. Laenen, G. Oderda and G. Sterman, \plb{438}{1998}{173},
   [arXiv:hep-ph/9806467]; E. Laenen, G. Sterman and W. Vogelsang,
   \prd{63}{2001}{114018} [arXiv:hep-ph/0010080]; A. Kulesza, G. 
Sterman and W. Vogelsang,
   \prd{66}{2002}{014011} [arXiv:hep-ph/0202251]; J. H. Kuhn, S. Moch,
   A. A. Penin and V. A. Smirnov, \npb{616}{2001}{286} [arXiv:hep-ph/0106298]
\bibitem{deFlorian1} D. de Florian and M. Grazzini, \npb{616}{2001}{247},
   [arXiv:hep-ph/0108273]
\bibitem{catanitwo} S. Catani, \plb{427}{1998}{161} [arXiv:hep-ph/9802439]
\bibitem{KOS} N. Kidonakis, G. Oderda and G. Sterman,
    \npb{531}{1998}{365} [arXiv:hep-ph/9803241]; {\it ibid} 
\npb{525}{1998}{299},
   [arXiv:hep-ph/9801268]
\bibitem{Sen} A. Sen, \prd{28}{1983}{860}
\bibitem{Akh} R. Akhoury, \prd{19}{1979}{1250}
\bibitem{CTG} C. F. Berger, T. Kucs and G. Sterman,
   \prd{65}{2002}{094031} [arXiv:hep-ph/0110004]
\bibitem{magnea} L. Magnea, \npps{96}{2001}{84} [arXiv:hep-ph/0008311];
   L. Magnea, \npb{593}{2001}{269} [arXiv:hep-ph/0006255]
\bibitem{sudakovCDR} L. Magnea and G. Sterman, \prd{42}{1990}{4222}
\bibitem{evolsuda} J. C. Collins in {\it Perturbative QCD}, edited by
A. H. Mueller, Advanced Series on Directions in High Energy Physics, Vol. 5
(World Scientific, Singapore, 1989); A. H. Mueller, 
\prd{20}{1979}{2037}; J. C. Collins,
    \prd{22}{1980}{1478}; A. Sen, \prd{24}{1981}{3281}; G. P. Korchemsky and
    A. V. Radyushkin, \npb{283}{1987}{342}; G. P. Korchemsky
    \plb{220}{1989}{629}
\bibitem{suda1L} G. Altarelli, R. K. Ellis and G. Martinelli,
    \npb{157}{1979}{461}; J. Kubar-Andre and F. E. Paige, \prd{19}{1979}{221};
    K. Harada, T. Kaneko and N. Sakai, \npb{155}{1979}{169}
\bibitem{suda2L} T. Matsuura, S. C. van der Marck and W. L. van
Neerven; \npb{319}{1989}{570} R. J. Gonsalves, \prd{28}{1983}{1542}
; G. Kramer and B. Lampe, \zpc{34}{1987}{497}; R. J. Gonsalves,
\prd{28}{1983}{1542}
\bibitem{factorjets} J. C. Collins and D. E. Soper, \npb{193}{1981}{381};
   G. Sterman, \npb{281}{1987}{310}; S. Catani and L. Trentadue,
   \npb{327}{1989}{323} {\it ibid} \npb{353}{1991}{183}
\bibitem{gammagq} S. Catani, E. D'Emilio and L. Trentadue,
   \plb{211}{1988}{335}; J. Kodaira and L. Trentadue, \plb{112}{1982}{66}
\bibitem{a3} C. F. Berger [arXiv:hep-ph/0209107]; S. Moch, J. A. M. 
Vermaseren and A. Vogt [arXiv:hep-ph/0209100]
\end{thebibliography}
\end{document}